# Reconstruction of Fine Scale Auroral Dynamics

Michael Hirsch, *Member, IEEE,* Joshua Semeter, *Senior Member, IEEE,* Matthew Zettergren, Hanna Dahlgren, Chhavi Goenka, *Member, IEEE,* Hassanali Akbari, *Member, IEEE*

*Abstract*—We present a feasibility study for a high frame rate, short baseline auroral tomographic imaging system useful for estimating parametric variations in the precipitating electron number flux spectrum of dynamic auroral events. Of particular interest are auroral substorms, characterized by spatial variations of order 100 m and temporal variations of order 10 ms. These scales are thought to be produced by dispersive Alfvén waves in the near-Earth magnetosphere. The auroral tomography system characterized in this paper reconstructs the auroral volume emission rate to estimate the characteristic energy and location in the direction perpendicular to the geomagnetic field of peak electron precipitation flux using a distributed network of precisely synchronized ground-based cameras. As the observing baseline decreases, the tomographic inverse problem becomes highly ill-conditioned; as the sampling rate increases, the signal-to-noise ratio degrades and synchronization requirements become increasingly critical. Our approach to these challenges uses a physics-based auroral model to regularize the poorly-observed vertical dimension. Specifically, the vertical dimension is expanded in a low-dimensional basis consisting of eigenprofiles computed over the range of expected energies in the precipitating electron flux, while the horizontal dimension retains a standard orthogonal pixel basis. Simulation results show typical characteristic energy estimation error less than 30% for a 3 km baseline achievable within the confines of the Poker Flat Research Range, using GPS-synchronized Electron Multiplying CCD cameras with broad-band BG3 optical filters that pass prompt auroral emissions.

## I. Introduction

Studies of the aurora using two or more cameras with overlapping fields of view (FOV) have been carried out for over a century [1], with more recent work focusing on the formal application of tomographic techniques [2]–[5]. Auroral tomography provides a means of accessing time-dependent information about remote auroral acceleration processes. In this technique, common volume measurements of the aurora from multiple ground-based imagers are used to reconstruct the wavelength-dependent ionospheric volume emission rate. The volume emission rate depends on the energy flux distribution of the precipitating magnetospheric electrons that have undergone a particular acceleration process, gaining high enough energy to penetrate deep into the ionosphere, giving rise to the auroral emissions via collisional and kinetic interactions with neutral species and ions. The volume emission rate reconstruction can be used together with a physics-based model of precipitating magnetospheric electrons to estimate the spatial distribution and characteristic energy of the primary electron differential number flux. Estimation and measurements of the precipitation characteristic energy have been used [6], [7] as a conduit to understand mechanisms driving auroral morphology at the finest spatio-temporal scales.

The reconstruction problem is challenging owing to uncertainties in model assumptions and the solution non-uniqueness that arises from the constrained viewing geometry. The use of a first-principles based physics model was motivated in part by the limited observation in the direction along the geomagnetic field $B_\parallel$. The short distance between cameras was motivated by the desire to get the highest feasible resolution in the direction perpendicular to the geomagnetic field $B_\perp$ [8]. These data inversion techniques provide the first realizable method of obtaining a persistent two-dimensional (energy, $B_\perp$) high resolution morphology estimate of the rapidly evolving electron precipitation above the ionosphere at the smallest ground-observable scales.

Auroral morphologies can be described in a Cartesian coordinate system, with axis $B_\parallel$ oriented along the Earth's local magnetic field $B$. Near Poker Flat Research Range, the inclination of the magnetic field is 77.5°, so the $B_\parallel$ axis is tipped 12.5° from the local geographic vertical axis toward magnetic south. The $B_\perp$ axis is defined to be orthogonal to $B_\parallel$ and coplanar with the cameras in this study. In auroral literature the "width" of auroral features refers to extent in the $B_\perp$ direction, and we follow this convention.

Prior work in auroral tomography [5], [9], [10] has focused almost exclusively on mesoscale features of $10^4$ m width recorded with typical sampling periods of order 1-30 seconds, with sensor baselines of 50-150 km. The peak auroral emission intensity typically lies in the altitude range of approximately 100-300 km. The $B_\parallel$ profile of the arc is dependent on the electron beam differential number flux and the characteristics of the neutrals and ions with which the precipitating particles interact. An active auroral display embodies a vast hierarchy of spatial scales. The global auroral oval is of order $10^5$ m width as measured along magnetic latitude from the poleward to equatorward edges. Dynamic fine-scale features embedded in an auroral breakup of $10^2$ m width are typically observed during the substorm expansion phase [11]. Anthropogenic aurora of $10^2$ m width has been observed from HAARP stimulus [12], [13]. A complete theory of the aurora must account for variations at all scales inherent in the phenomena. Although our theoretical understanding of global and mesoscale variability, and its drivers in the solar wind and magnetosphere, is well developed [14], the physics underlying decameter-scale structure embedded within active auroral displays remains incomplete.

M. Hirsch is with the Department of Electrical and Computer Engineering, Boston University, Boston, MA, 02215 USA e-mail: mhirsch@bu.edu.

Thanks to H. Murato and G. Thayer with the Boston University Scientific Instrument Facility for mechanical design and build aspects of this system.

Thanks to A. Baurley for assisting in simulation runs and analysis.

Source code available at https://github.com/scienceopen/histfeas

This work was funded by the National Science Foundation Atmosphere and Geospace Science Directorate under Grant 1216530 and Grant 1237376.

Manuscript received ; revised .

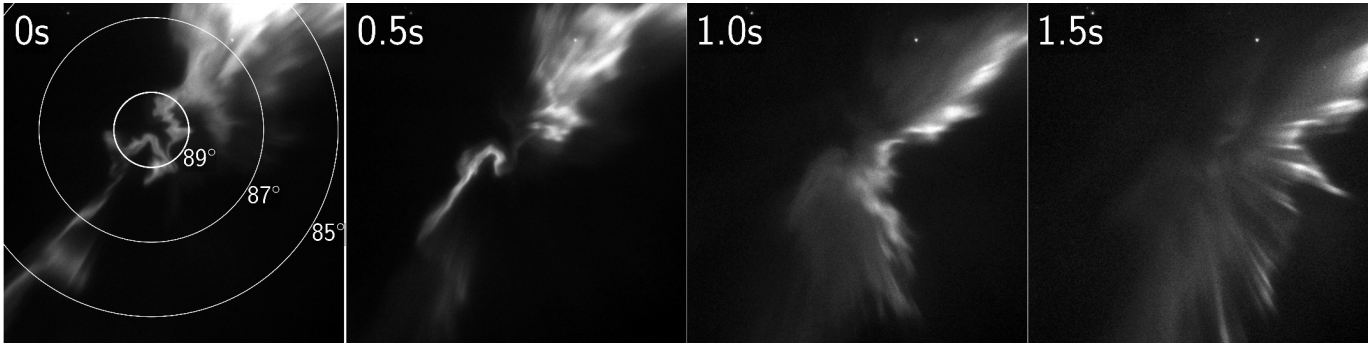

Fig. 1. Radical change in perspective for 100 m structure in 1.5 seconds due to apparent $B_\perp$ transverse motion [8], [11]. Contours are centered on local magnetic zenith.

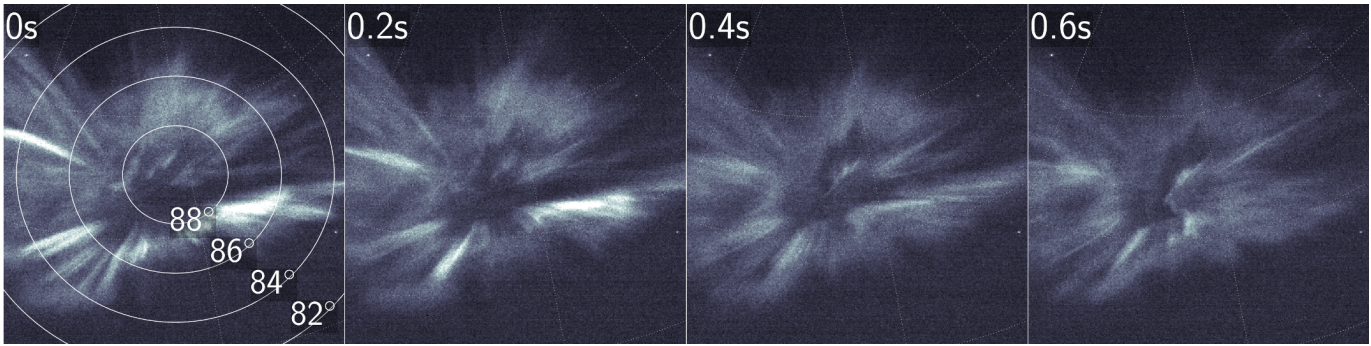

Fig. 2. Flaming aurora evolution over 600 ms [18]. Contours are centered on local magnetic zenith.

Auroral structures of sub-100 m width have been known to exist for decades [15], [16], and are seen regularly in long-term observations with modern cameras. An example of a thin 100 m wide auroral structure exhibiting rapid lateral motion is shown in the image sequence of Fig. 1. Note the substantial change in appearance of the arc in 1.5 seconds, corresponding to a 5° change in observer perspective, or about 10 km in the $B_\perp$ dimension assuming 120 km apparent auroral feature altitude. An example of flaming aurora [17], [18] evolving over 600 ms is shown in Fig. 2.

The tomographic techniques applied in this paper will contribute to our understanding of how such ephemeral fine-scale structure emerges in the incoming particle flux. The observational requirements for a tomographic imaging system capable of resolving these scales are extreme, and the resulting inverse problem is highly ill-conditioned. This paper presents a feasibility study for a high frame-rate, short baseline, auroral tomography system we refer to as the High Speed Tomography system, or HiST [19]. Through simulation and modeling, we demonstrate that Electron Multiplying CCD (EMCCD) camera technology coupled with a physics-based regularization scheme is capable of resolving electron differential number flux dynamics of order 100 m and 10 ms.

The remainder of the paper is organized as follows: Section II describes the geometry leading to observations with a modeled camera setup fitting within the confines of the Poker Flat Research Range. Section III describes the forward model employed in the design of the camera system and analysis of the system data. Section IV describes the data inversion process, with a summary of the methods used. Section V presents estimates of the differential number flux obtained via inversion of forward model observations, along with the volume emission rate and camera pixel intensities that would result from the estimated differential number flux. Section V-A gives forward and inverse simulated results for aurora with an apparent translation along $B_\perp$. Section V-B gives forward and inverse simulated results for a flaming aurora scenario. Section VI gives the conclusions drawn from the simulations carried out with the model and data inversion process described herein.

## II. OBSERVATIONAL REQUIREMENTS

A key observational goal of the HiST system is to increase resolution in the horizontal ($B_\perp$) dimension to the finest physical scale that is tractable given imaging geometry and inversion technique limitations. We wish to perform a tomographic analysis of features that are 100 m wide in the altitude range of 100-300 km and with temporal variation of order 10 ms. The requisite tomographic imaging system must have camera baselines of order 1-10 km [8], high sample rates of 50-100 frames/s, precise ($\ll$ 1 ms) time synchronization and image angular registration.

Even if the observational requirements are met, a further challenge lies in the extreme ill-conditioning of the resulting inverse problem. A key ingredient in this study is a regularization scheme to handle the poorly observed vertical ($B_\parallel$) dimension. Fairly complete models exist for predicting the distribution of optical emissions along a magnetic field line given an incident energetic electron spectrum. This work leverages the TRANSCAR (from the French acronym for "transport squared" [20]) particle penetration model [21]. Other candidate models available include GLOW [22], Rees/Sergienko/Ivanov [23] as well as earlier models [24]. We chose TRANSCAR because of the simulation outputs immediately applicable to incorporating incoherent scatter radar and optical observations, as well as a first-principles 1-D time-dependent modeling scheme incorporating a fluid model with kinetic update step [25]. Using the TRANSCAR computed excitation rates to produce eigenprofiles for precipitating electron beams with energies in the 58 eV to 17.7 keV range [18], we develop and test an algorithm to estimate the differential number flux responsible for the ground-observed auroral intensity vs. angle.

Fig. 3 depicts a representative two-camera viewing geometry with a 3 km baseline in the Poker Flat Research Range. This geometry is chosen to be coplanar with each camera and a line extending toward the local magnetic zenith. The relatively narrow 9° field of view (FOV) and boresight angles depicted in Fig. 3 leads to a overlapping $B_\perp$ observable range on the order of 10 km. Choosing a wider FOV inherently leads to more viewing angles far from magnetic zenith, where the fine horizontal detail will be washed out [8]. The black grid of Fig. 3 represents the model $(B_\perp, B_\parallel)$ coordinate space.

This feasibility study evaluates whether a reliable parameterization of the differential number flux spectrum may be extracted from such a two-camera configuration. The first phase of HiST consists of a pair of sensitive EMCCD cameras synchronized using an ASIC (application-specific integrated circuit) with a GPS disciplined oscillator timebase, such that timing jitter intrinsic to internal camera electronics is the dominant source of timing error. The final system image timing error is much less than 1 ms as measured with a custom FPGA LED test device designed as part of this effort [19]. Geographic azimuth between sites is 286.4°, and the magnetic azimuth angle is 306.3° according to the IGRF 11 [26] model for April 2013. The HiST phase I two-site field deployment with a 3 km camera baseline as depicted [27] in Fig. 4 is at WGS84 coordinates $(65.119, -147.432)$ and $(65.127, -147.497)$, corresponding to the viewing geometry of Fig. 3. Each phase I HiST site consists of an Andor iXon EMCCD camera with 512 x 512 pixel image sensor and a lens providing a 9° FOV. The Schott BG3 filter suppresses non-prompt emissions with lifetimes covering many tens of frames. Each camera is pointed at local magnetic zenith, with star calibration and localized plate scaling accomplished via the Astrometry.net program [28]. A successful realization of the system modeled in this feasibility study will estimate the time-dependent differential electron number flux $\Phi(z, \mu, E)$ [cm$^{-2}$ s$^{-1}$ eV$^{-1}$ ster$^{-1}$] at the top of the ionosphere, designated $\Phi_{top}(E)$, where $z$ is the altitude along the $B_\parallel$ dimension, $\mu$ is the pitch angle and $E$ is the differential energy bin, assuming a fixed pitch angle distribution encapsulated in $\mu$. Since we are interested in Alfvénic aurora, we assume a field-aligned pitch angle distribution [8]

$$\Phi_{top}(E) = \Phi(z = 1000 \text{ km}, \mu = 0, E) \tag{1}$$

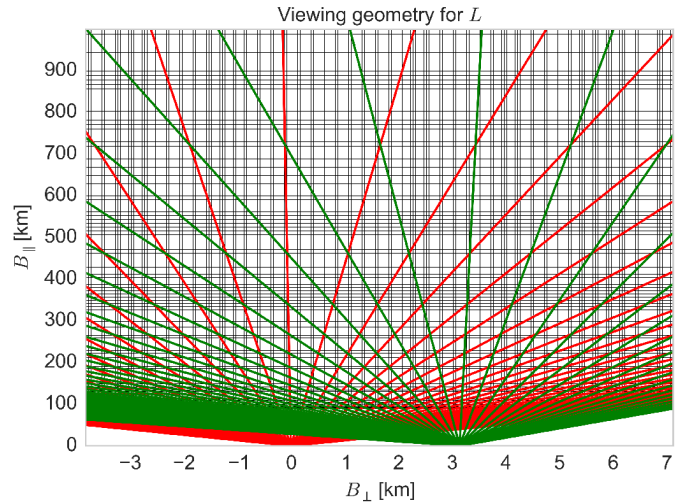

Fig. 3. Viewing geometry for the two-camera system at the Poker Flat Research Range, showing selected lines of sight over a decimated reconstruction grid.

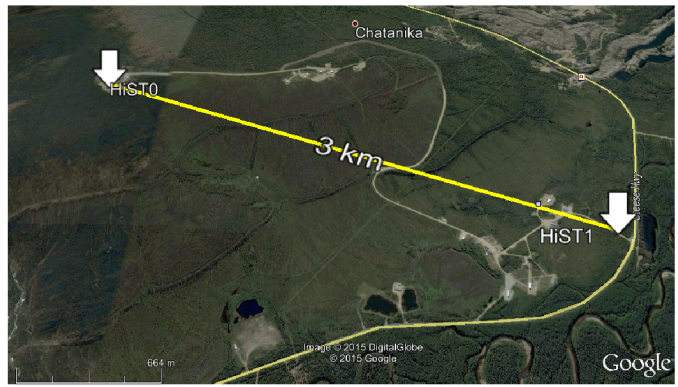

Fig. 4. Poker Flat Research Range, Alaska: HiST Phase I camera locations

In the remainder of this paper we take $\Phi_{top} \triangleq \Phi_{top}(B_\perp, E)$ implicitly as a function of energy and space in the $B_\perp$ direction.

### III. FORWARD MODEL

For the extremely ill-conditioned problem produced by cameras with 3 km spacing over 100 km from the target of interest, regularization of the poorly observed vertical dimension is a necessary step to get a tractable inversion in terms of computational effort and error bounds. We fashion a set of eigenprofiles from a problem comprised of linear differential equations by first making the assumption that production $p_{ij}$ and loss $l_{ij}$ for prompt emissions at steady state are related by

$$p_{ij} - l_{ij} = 0 \tag{2}$$

for the j$^{th}$ excitation process of the i$^{th}$ species. The eigenprofiles are used as basis functions [18] in a linear system tying together ground-observed auroral intensity to auroral volume emission rate via a known viewing geometry. We can solve for the coefficients that correspond to the differential number flux for each log-spaced energy bin, yielding estimated differential number flux $\hat{\Phi}_{top}(B_\perp, E)$ for each new set of camera images.

To model the excitation rates due to primary electron precipitation, we use the 1-D TRANSCAR model [18], [25], [29]–[31]. Primary considerations for use of TRANSCAR include that 192 spectra are derived [20] from the excitation rates modeled by TRANSCAR. The TRANSCAR hybrid kinetic/fluid time-dependent ionosphere model becomes more relevant in future studies incorporating joint observations with instruments such as incoherent scatter radar. The use of a large number of spectra is important to maximizing the information available from a broadband optical filter such as the BG3 that passes numerous prompt line emissions. Because a key requirement of the system is capturing order 10 ms auroral dynamics, it was desirable to capture and incorporate the largest number of spectra possible to increase SNR at high frame rates. TRANSCAR is a physics-based model of six positive ion species and their neutral parents: $O^+$, $H^+$, $N^+$, $N_2^+$, $NO^+$, $O_2^+$ along with electrons $e^-$ using the charge neutrality [21], [29], [30] of plasma $n_e = \sum_S n_s$. An 8-moment model [21] encompasses thermal diffusion effects so that important heat flows are captured [30]. The TRANSCAR excitation rates and eigenprofiles used in this feasibility study are computed once for a particular set of geophysical parameters in an offline manner, which takes about 30 minutes using the idle CPU cycles of office PCs arranged in a compute cluster via GNU Parallel [32]. The rest of the forward model is implemented in about 2 seconds. The data inversion that must be executed for each observation time step must be done on-line for each new observation and takes about one minute on a desktop PC, depending on the number of cells in the projection matrix $\mathbf{L}$.

The close-spaced optical instruments used in this study yield persistent observations of precipitation process outcomes [33], [34] complementing on-orbit and rocket-borne *in situ* measurements with a broader spatiotemporal context, along with improved $B_\perp$ resolution over widely spaced ground-based imagers. Observation of a typical rapidly moving (several km/s) auroral feature implicitly requires a frame rate on the order of 100 Hz for a narrow 9° FOV and megapixel-class imager. Cameras comprising a multi-camera tomography system must have their frame start/end exposure times known to better than 1/10th of a single frame, or a data inversion will have limited scientific utility since the emissions observed at time $t_0$ at HiST0 will be smeared together with the results at time $t_0 + \varepsilon$ at HiST1 due to timing error $\varepsilon$. The camera site spacing is chosen based on the forward model described in this section along with practical facility availability.

The auroral target of interest is taken to operate within the following first-order constraints:

1) Auroral behavior in the $B_\parallel$ dimension is strongly influenced by time-dynamic electron particle penetration [29], as modeled by TRANSCAR. Time of flight difference between high energy and low energy particles in the lower magnetosphere at time scales less than order 10 ms have been observed [35]. The tomographic process gives information on vertical structure not available in zenith-oriented line integrations alone as in [35], so our technique will capture dynamics with frame rates to at least 100 Hz.
2) Precipitating $e^-$ acceleration has taken place above the uppermost altitude cell of the 1-D model, implying that thermospheric and mirroring forces are neglected [29], [36]
3) Auroral behavior in the $B_\perp$ dimension is dominated by collisionless processes above the "top" of the ionosphere (altitude > 1000 km) [37], [38]

With these constraints in mind, we continue with a discussion of the quantitative particulars of the models and algorithms used in this feasibility study.

Referring to the left column of Fig. 5, the forward model input $\Phi_{top}$ is generated using a parameterization [39] with representative values shown in Fig. 6, where the location in energy of peak differential number flux is known as the characteristic energy $E_0$. The physical process generating $p_\lambda(z)$ in the second block of the left column of Fig. 5 is modeled in TRANSCAR [21], [29], [30] and represented by the eigenprofiles **T** in the upper right block of Fig. 5, with line-integrated modeled spectra for each beam energy shown with and without BG3 filtering in Fig. 7. Some of the brightest features in the aurora are produced by metastable transitions with radiative lifetimes of order 1-10 s [40]. In Alfvénic aurora, the electron flux rapidly changes ($< 10$ ms scales) in $B_\perp$ and $E_0$, and the intense metastable emissions glow like an high-persistence oscilloscope phosphor, which in a white light sensor can cover up the much fainter prompt emissions that have several orders of magnitude shorter lifetimes. Each camera was equipped with a BG3 optical filter with the transmission characteristics of Fig. 8 to greatly attenuate these long lifetime features. In particular, the deep notch in transmission for the long lifetime metastable emissions lines includes 557.7 nm and 630 nm. The volume production rate of process $p_{ij}$ integrated over fixed pitch angle $\mu$ resulting from the TRANSCAR model is [8],

$$p_{ij}(z) = n_i(z) \int \sigma_{ij}(E) \Phi(z,E) dE \quad (3)$$

where $n_i$ is the MSIS90-initialized density of the $i^{th}$ ground-state neutral species (e.g. $N_2$, $O_2$, O). $\sigma_{ij}$ is the electron impact cross section of the $j^{th}$ excitation process for the $i^{th}$ species. $\Phi(z,E)$ is the pitch angle integrated flux obtained from the 1-D model TRANSCAR [21] for 33 log-spaced energy bins $E$ ranging from 58 eV to 17.7 keV [18]. For prompt emissions, we connect excitation rates to optical volume emission rates using (2) with [20], [40] the Einstein coefficients and Franck-Condon factors,

$$p_\lambda(z) \propto p_{ij}(z) \quad (4)$$

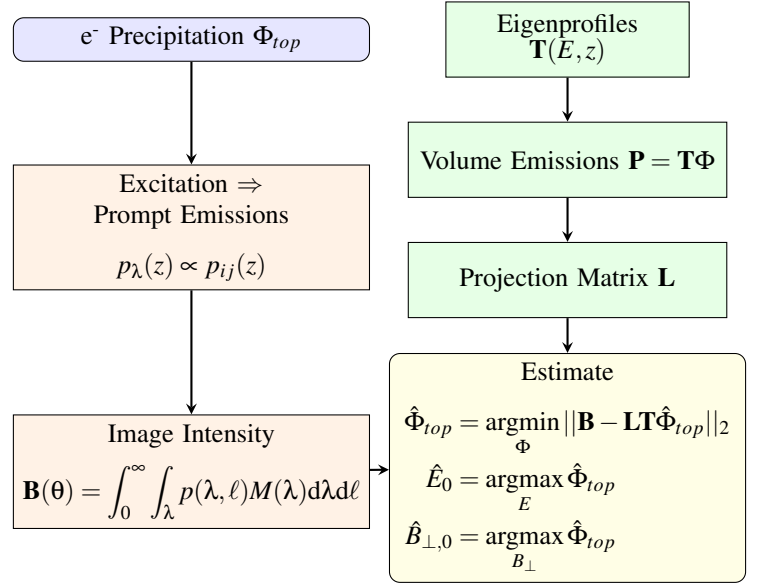

Fig. 5. Block diagram of HiST auroral tomography forward model and data inversion.

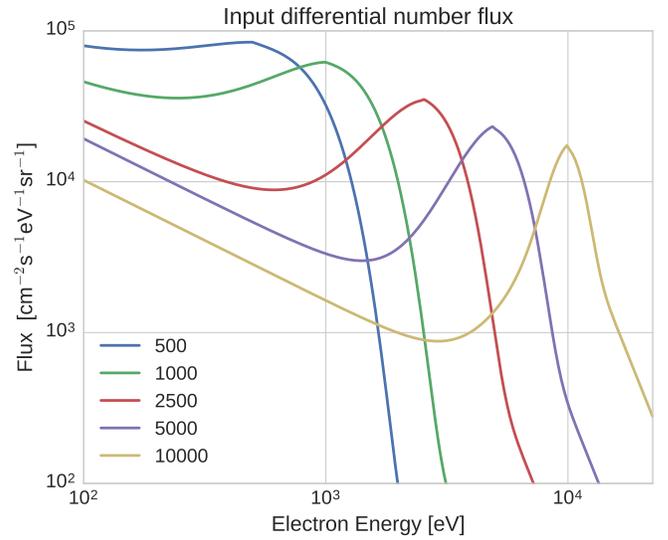

Fig. 6. Input differential number flux for beams with $E_0 \in \{500, 1000, 2500, 5000, 10000\}$ eV [39].

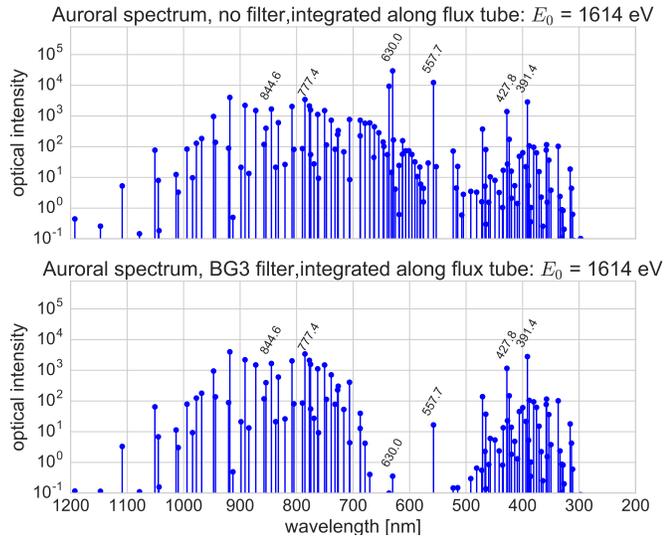

Fig. 7. Auroral Spectrum integrated along flux tube for $E_0 = 1.6$ keV, with and without BG3 filter.

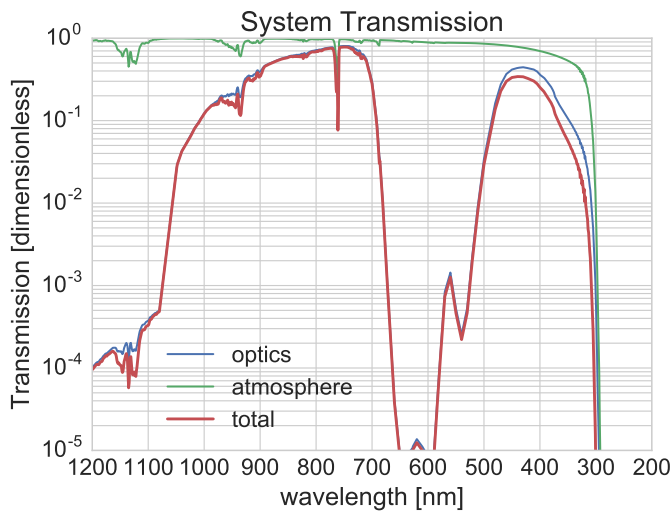

Fig. 8. Optical system transmission, including BG3 filter, EMCCD window and LOWTRAN modeled atmospheric absorption.

For the lower left block of Fig. 5, the photon flux at the $k^{th}$ camera pixel is described by a line integral mapped via the lens to angle $\theta_k$, treating the auroral region as optically thin at the wavelengths observable through the optical filtering and LOWTRAN [41] modeled atmospheric absorption of Fig. 8. Considering (4) and total transmission $M(\lambda)$ shown in Fig. 8, the camera photon flux $\mathbf{B}(\theta)$ is

$$\mathbf{B}(\theta) = \int_0^\infty \int_\lambda p(\lambda, \ell) M(\lambda) d\lambda d\ell \quad (5)$$

The camera exposure time $\tau$, amplifier gain $g$ and pixel area $a$ are modeled with the output in data numbers **D** as:

$$\mathbf{D} = \tau a g \mathbf{B} \quad (6)$$

where typical values include $a = (16 \mu\text{m})^2, \tau = 2 \times 10^{-2}$ s, $g = 1$ $D/e^-$.

Referring to the right column of Fig. 5 we assemble projection matrix **L** by mapping viewing angle $\theta$ to our discrete EMCCD imaging arrays, and compute the intersection length of each ray [42] with the relevant cell of **L** using the Cohen-Sutherland line clipping algorithm [43]. The dimensions of **L** are $N_{cam}N_{cut} \times N_{B_\perp}N_{B_\parallel}$, where $N_{cam}$ is the number of cameras in the system, $N_{cut}$ is the number of 1-D pixels used from each camera and $N_{B_\perp}, N_{B_\parallel}$ are the number of $B_\perp, B_\parallel$ pixels in the grid for the volume emission rate matrix **P**. The IGRF 11 model is incorporated into **L** for the Poker Flat Research Range, where the inclination 77.5° and declination 19.9° of the local geomagnetic field determine the angular coordinates of magnetic zenith.

The grid of Fig. 3 extends from approximately 90-1000 km altitude, showing the locations used in estimating volume emission rate **P** due to the incident differential number flux $\Phi_{top}$. Overlaid on this grid are the decimated 1-D rays corresponding to intensity vector $\mathbf{B}(\theta)$. For Fig. 3 and the analysis of Section V-A and V-B, $N_{cam} = 2, N_{cut} = 512, N_{B_\perp} = 219, N_{B_\parallel} = 123$. This forward model yields ground-observed optical intensity vs. angle due to electron differential number flux $\Phi_{top}(B_\perp, E)$. The analysis in Section IV uses observations from ground-based cameras to estimate the unobservable differential number flux $\hat{\Phi}_{top}$ via a minimization algorithm.

## IV. Data Inversion

To estimate the characteristics of the time-dependent differential electron number flux $\Phi_{top}$ high in the ionosphere where collisionless processes dominate we employ a physics-based regularization scheme. The poorly observed $B_\parallel$ dimension is regularized with a linear basis expansion of volume emission rate eigenprofiles calculated by the TRANSCAR model. The 33 log-spaced energy bins from 58 eV to 17.7 keV each have a coefficient estimated by our inversion algorithm for each $B_\perp$ location, comprising $\hat{\Phi}_{top}(B_\perp, E)$. For the simulations of Sec. V, $\hat{\Phi}_{top}$ has dimensions $(219, 33)$. Regularization along $B_\parallel$ is key to finding a physically plausible solution from the infinitely many possible solutions due to the large null space of the inverse problem. The data inversion process is outlined in the right column of Fig. 5. As observed in the middle row of Fig. 11, the volume emission rate is a smooth function of differential number flux and altitude. The smoothness justifies describing the relationship between the unobservable *in situ* physics and the observable auroral intensity by the Fredholm Integral of the First Kind,

$$g(s) = \int_a^b K(s,t) f(t) \mathrm{d}t \qquad (7)$$

where $f(t)$ is the unknown quantity, $g(s)$ is the observed quantity, and $K(s,t)$ is the kernel through which $g(s)$ is observed, encompassing optical filters, line integration of volume emission rate, and noise. For the present auroral tomography problem, we incorporate TRANSCAR eigenprofiles

$$T(E,z) = \int_\lambda p_\lambda(E,z) M(\lambda) \mathrm{d}\lambda \qquad (8)$$

in a representation of total auroral volume emission rate as

$$P(z) = \int_0^\infty T(E,z) \Phi_{top}(E) \mathrm{d}E \qquad (9)$$

which has the same form as (7) and may be discretized in matrix form,

$$\mathbf{P} = \mathbf{T}\Phi_{top} \qquad (10)$$

The discretized forms are convenient for computer implementation since the continuous integration (9) is represented by matrix multiplication (10). The BG3 filtered and atmosphere attenuated continuum of wavelengths is observed at the camera as grayscale intensity

$$\mathbf{LT}\Phi_{top} = \mathbf{B} \qquad (11)$$

resulting in the data numbers $\mathbf{D}$ of (6).

In general $\mathbf{L}$ and $\mathbf{T}$ are not square, so the inverse $\mathbf{L}^{-1}$ and $\mathbf{T}^{-1}$ are not defined in this underdetermined system. The ill-conditioned and Hadamard ill-posed nature of the problem arises both from the extreme problem geometry and that there is not a unique tomographic solution for the incident number flux causing an auroral display. A method for solving such problems via brute force is the use of minimization algorithms [44]. The algorithm selected for this effort is the Limited Memory Broyden-Fletcher-Goldfarb-Shanno algorithm [45]–[47], known as L-BFGS-B [48]. This algorithm was selected based on fast convergence for the very large number ($> 7000$) of $\hat{\Phi}_{top}(B_\perp, E)$ parameters to minimize based on an empirical comparison with other contemporary minimization techniques.

For a particular realization of geophysical conditions and choice of differential number flux energy bins $\mathbf{T}$ is obtained from an off-line computation. As implicit in (7), $\mathbf{T}$ is identical in the forward model and data inversion. The 1-D slices of the synchronized images from each camera are stacked in column-major vector $\mathbf{B}$. The 2-D array $\hat{\Phi}_{top}(B_\perp, E)$ has rows arranged by precipitation energy in eV and columns arranged by $B_\perp$ location in kilometers. We use L-BFGS-B minimization function

$$\hat{\Phi}_{top}(B_\perp, E) = \operatorname*{argmin}_\Phi ||\mathbf{B} - \mathbf{LT}\hat{\Phi}_{top}||_2 = \operatorname*{argmin}_\Phi ||\mathbf{B} - \hat{\mathbf{B}}||_2 \qquad (12)$$

with the bounds $\Phi \in [0, \infty)$ is given an initial guess $\Phi_{top}(B_\perp, E) \equiv 0$, and is allowed to run for 10–20 cycles. Automated measurements of $E_0$ and $B_{\perp,0}$ are accomplished via a 2-D Gaussian fitter algorithm originally based on MINPACK [49]. The region of the maximum differential number flux is fitted with a 2-D Gaussian using a Levenberg-Marquardt least squares algorithm to find the parameters best fitting the peak vicinity of $\hat{\Phi}_{top}$. In general, the forward model will have limitations in absolute accuracy with regard to the physical world due to the model assumptions and simplifications necessary to get a tractable computation within time and memory constraints. We now examine simulations of two types of highly dynamic auroral events to show the feasibility of the HiST system for estimating $E_0$ and $B_{\perp,0}$.

## V. Simulations

This feasibility study includes two types of auroral morphology simulations: horizontally translating aurora and flaming aurora. Horizontally translating aurora may have up to several km/s $B_\perp$ motion during substorms. Flaming aurora [17] morphology is categorized as an apparent rapid increase in altitude of the auroral peak emission–within less than a second [18]. In both simulations, we use the BG3 filter transmission of Fig. 8. The arcs have been modulated with a Gaussian shape yielding a $B_\perp$ width of about 100 m. We have added Poisson distributed noise $\varepsilon$ to the observed optical intensity vector $\mathbf{B}$. Throughout this analysis $\mathbf{B} \equiv \mathbf{B} + \varepsilon$ where

$$\varepsilon(k; \lambda_p) = \frac{\lambda_p^k \exp(-\lambda)}{k!} \qquad (13)$$

for each forward modeled intensity vector to simulate the noise inherent in realizable systems. The simulation time step runs typically 10 times faster than the simulated exposure. For the simulations presented the simulation time step was 2 ms while the simulated camera exposure was 20 ms. As in the real world system, the rapid motion of the simulated aurora spatially smears the observed intensity since the exposure time of 20 ms is long compared to the auroral temporal dynamics.

Savitzky-Golay filtering [50] of order 2 and support width 15 is used for both simulations to reduce the impact of the observation noise, as would be apparent with real camera data. We use two cameras with locations $B_\perp \in \{0, 3\}$ km taken as representative of the camera spacing achievable within the confines of the Poker Flat Research Range. This simulation covers 10.9 km along $B_\perp$ (horizontal), with the $B_\perp$ cell size set to 50 m. Along the $B_\parallel$ (altitude) dimension a smoothly varying grid size is used, with the finest grid spacing at low altitudes to capture the dynamics of the auroral peak emission region. At higher altitudes the coarser grid spacing saves computational time and memory. This $B_\parallel$ grid configuration is cumulatively defined at each step $dB_\parallel$:

$$dB_\parallel = \tanh(\tau), \quad \tau \in [0, 3.14], \quad B_\parallel \in [90, 1000] \quad (14)$$

The L-BFGS-B algorithm rapidly converges for several steps before making a very slow approach to the $\hat{\Phi}_{top}$ estimate in light of noise in **B** and perturbations in **L**, so we typically truncate the minimization after 10–20 iterations. For convenience we denote the estimated peak location in energy and space of the precipitation differential number flux

$$\hat{\Phi}_{top,0} \triangleq \hat{\Phi}_{top}(B_{\perp,0}, E_0) \quad (15)$$

### A. Model and Inversion of Laterally Translating Aurora

The laterally translating aurora simulation uses $E_0 \equiv 5$ keV and $B_{\perp,0} \in \{1.55, 3.75\}$ km. Fig. 9(a)(c) show the input differential number flux $\Phi_{top}$, resulting in the volume emission rate **P** displayed in Fig. 9(e)(g). Fig. 9(b)(d) shows the estimated differential number flux $\hat{\Phi}_{top}$ using the L-BFGS-B algorithm and two cameras at $B_\perp \in \{0, 3\}$ km. Table I describes the estimated differential number flux results. The artifacts in $\hat{\Phi}_{top}$ and $\hat{\mathbf{P}}$ come from the noise deliberately injected into the simulated **B**. These artifacts are smaller in amplitude than the peak closest to the true answer, allowing for $\hat{\Phi}_{top,0}$ to be extracted despite the artifacts. The estimated volume emission rate $\hat{\mathbf{P}}$ shown in Fig. 9(f)(h) has morphologically similar characteristics to the forward modeled volume emission rate in Fig. 9(e)(g), as expected.

### B. Model and Inversion of Flaming Aurora

We model two time steps of a flaming auroral event with $E_0 \in \{4.5, 1.6\}$ keV. The input differential number flux $\Phi_{top}$ is shown in Fig. 10(a)(c). Fig. 10(b)(d) show the estimated differential number flux $\hat{\Phi}_{top}$ using the L-BFGS-B algorithm. 1-D cuts of $\Phi_{top}$ and $\hat{\Phi}_{top}$ at $B_\perp = 1.0$ km are shown in Fig. 11(a)(b) respectively to aid in visualizing the characteristic energy $E_0$. Table I describes the $E_0$ estimation error.

The estimated volume emission rate $\hat{\mathbf{P}}$ as shown in Fig. 10(f)(h) and as 1-D cuts in Fig. 11(c)(d) have morphologically similar characteristics to the forward modeled volume emission rate **P** in Fig. 11(e)(g), as expected. We observe that the estimated ground-observed intensity $\hat{\mathbf{B}}$ in Fig. 11(e)(f) is within a small factor of the forward model intensity **B**. As summarized in Table I, $\hat{\Phi}_{top,0}$ has been estimated with $E_0$ error typically less than 30% for simulated auroral arcs within 2.5° of magnetic zenith.

The addition of a third camera at $B_\perp = 10$ km initially does not appear to make a dramatic improvement in increasing the angular range from magnetic zenith for estimating $\hat{\Phi}_{top,0}$. It is desirable to extend the estimate of precipitation characteristics to 3-D, which intrinsically motivates incorporating more than 2 cameras into HiST. As observed in Table I, the $\hat{\Phi}_{top,0}$ estimate is usable to at least 2.5° from magnetic zenith. It is apparent that a key limit of the $B_\perp$ range of the inversion is keeping the auroral target within the common FOV of the cameras, as is trivially expected.

## VI. CONCLUSIONS

In this paper, we have shown results from a regularization scheme using the physics encapsulated in TRANSCAR modeled eigenprofiles in a two camera simulation, with testing extended to three cameras for future 3-D work. We observed estimates of the peak differential number flux $\hat{\Phi}_{top,0}(B_{\perp,0}, E_0)$ for an auroral arc in the common FOV of the cameras, with typical error less than 30% for auroral arcs within 2.5° from camera boresight on magnetic zenith. TRANSCAR is used in a linear basis expansion of log-spaced energy bins across an energy range observed in the most common auroral events, enabling future extension to incorporate incoherent scatter radar and other instruments to form a meta-instrument for observing the ionospheric short term and long term trends. This basis expansion is used to regularize the poorly observed vertical dimension, simultaneously enabling high spatial resolution in $B_\perp$, which is important for capturing the detail in extremely dynamic dispersive auroral events with 10 ms temporal scales. The performance estimates of this feasibility study show that a two camera system at the Poker Flat Research Range with 3 km camera separation can give new science insights on multiple fronts, including the highly dynamic electron beam structures driving into the ionosphere. Specifically, we can estimate the characteristic energy and $B_\perp$ peak location of the differential number flux $\Phi_{top}$. The new observation techniques include use of filtered broadband optical emissions to select only prompt emissions with fast, highly sensitive EMCCD cameras, enabling the use of high frame rates with cadence of order 10 ms. The modeled HiST instrument is shown to be capable of high resolution electron precipitation characteristic energy estimates along $B_\perp$ within suitable error bounds, while retaining the qualitative morphology of the differential number flux in the spatial and energy domains. Future work includes extending this estimate to 3-D by utilizing 3-D phantoms in the forward model and 3-D inversion of the 2-D pixel intensity images from the cameras, along with a 3 camera phase II HiST deployment to Poker Flat Research Range for a multi-year autonomous deployment beginning in the 2016 auroral season.

*Index Terms*—aurora, optical tomography, ionosphere, remote sensing

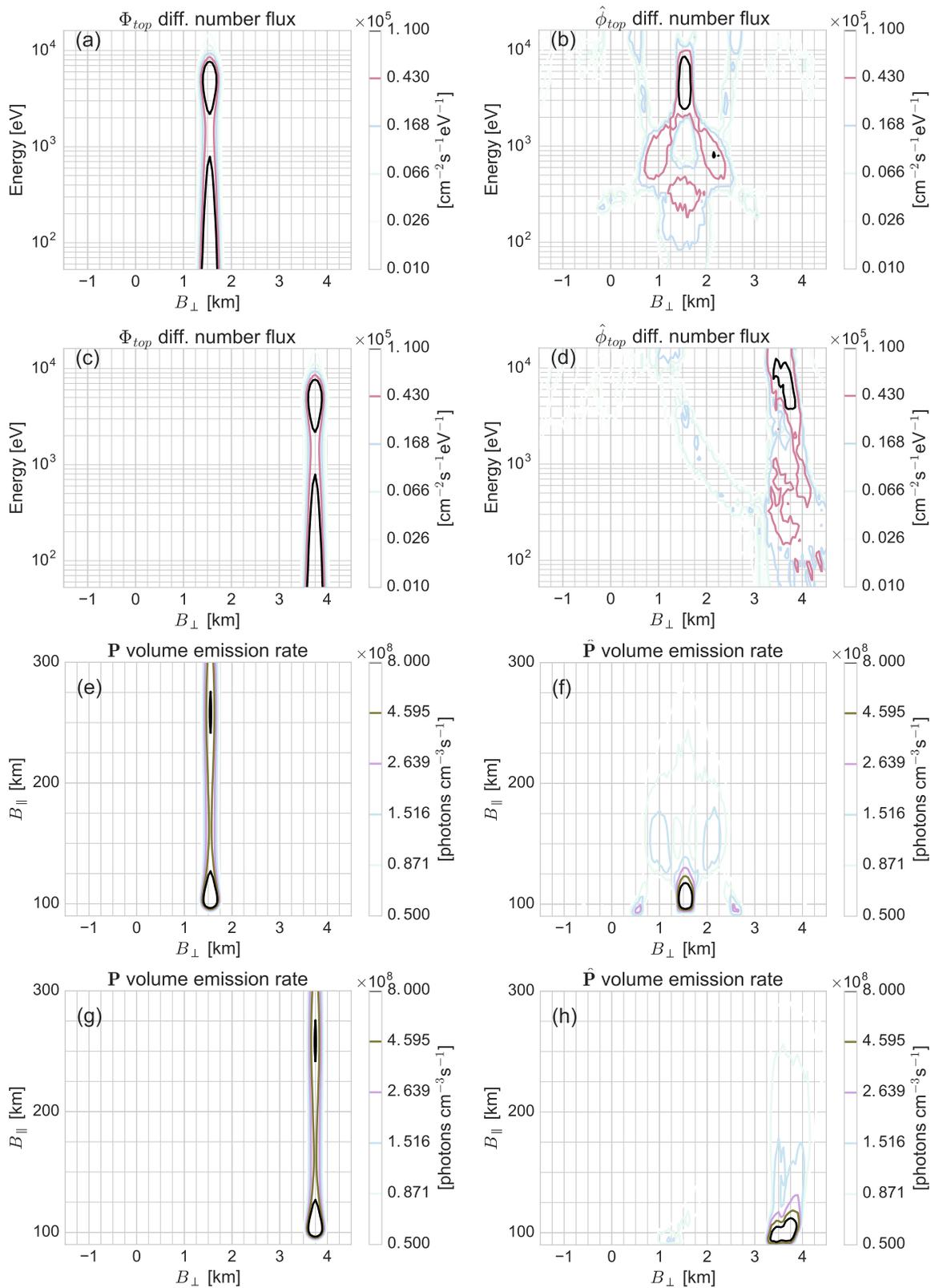

Fig. 9. $B_\perp$ translating aurora simulation with characteristic energy $E_0 \equiv 5$ keV and $B_{\perp,0} \in \{1.55, 3.75\}$ km. (a)(c): differential number flux $\Phi_{top}$. (b)(d): estimated differential number flux $\hat{\Phi}_{top}$. (e)(g): forward modeled volume emission rate $\mathbf{P}$. (f)(h): estimated volume emission rate $\hat{\mathbf{P}}$.

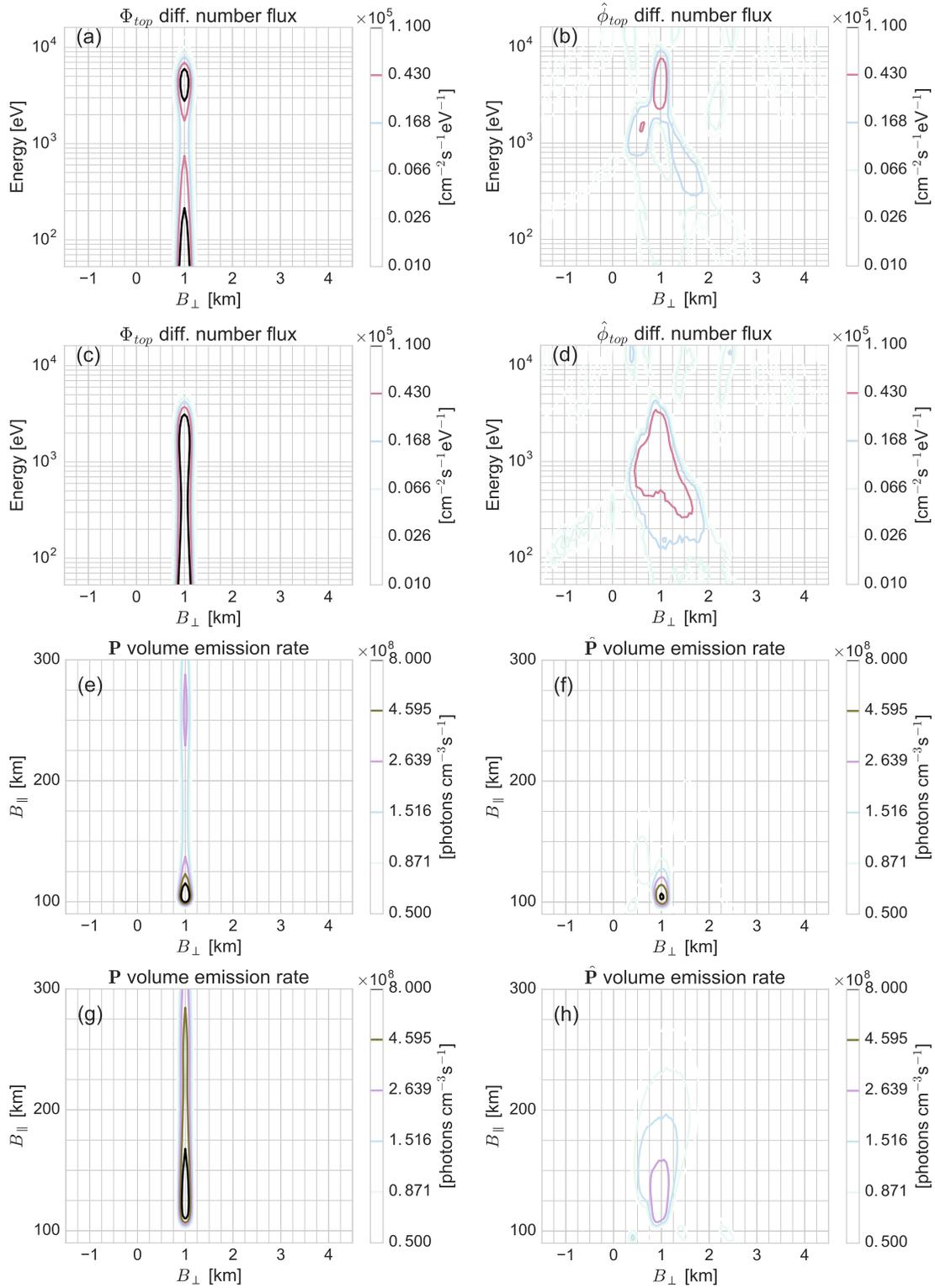

Fig. 10. Flaming aurora simulation with characteristic energy $E_0 \in \{4.5, 1.6\}$ keV and $B_{\perp,0} = 1.0$ km. (a)(c): differential number flux $\Phi_{top}$. (b)(d): estimated differential number flux $\hat{\phi}_{top}$. (e)(g): forward modeled volume emission rate $\mathbf{P}$. (f)(h): estimated volume emission rate $\hat{\mathbf{P}}$.

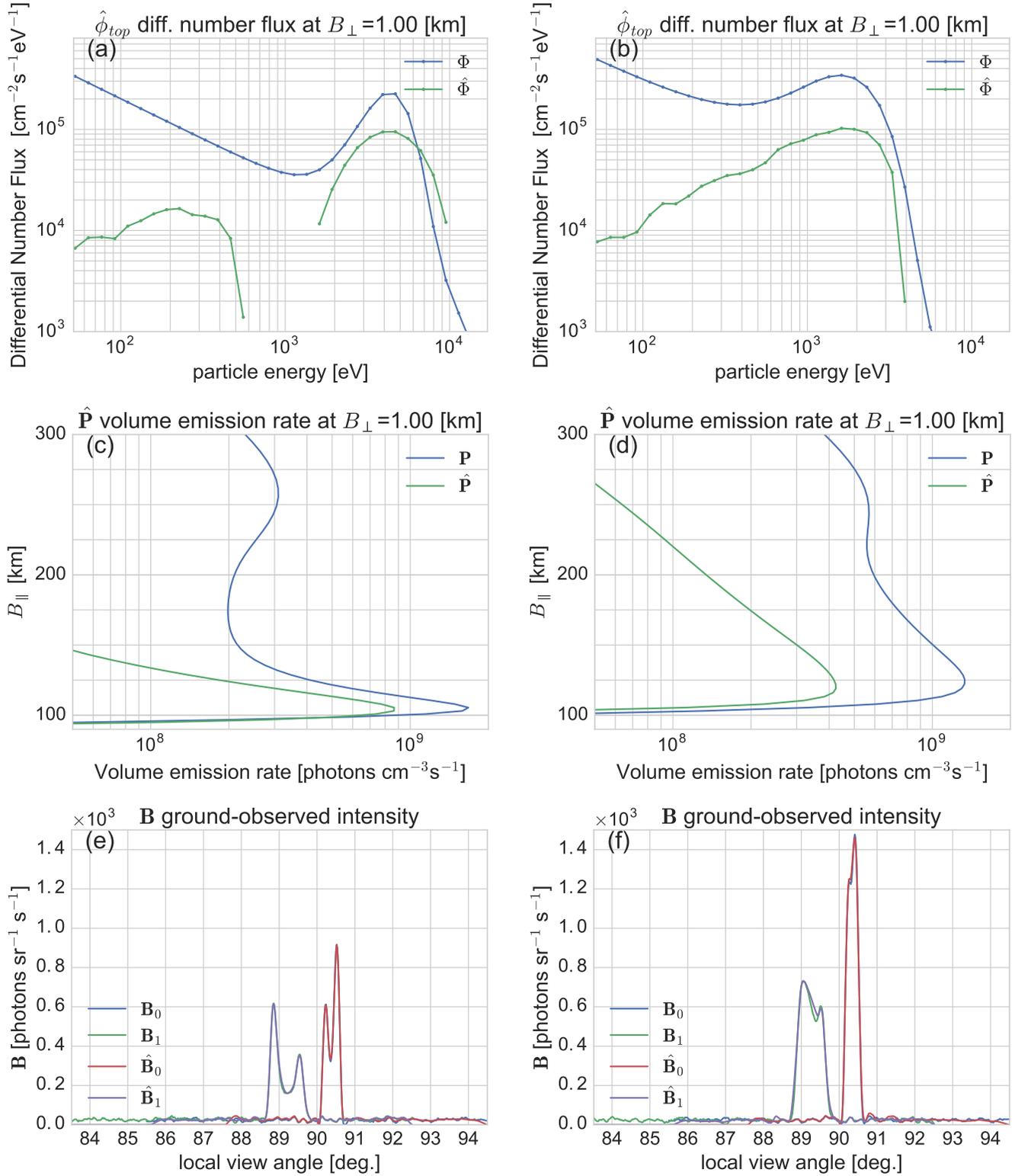

Fig. 11. Flaming aurora simulation, 1-D cuts. (a)(b): Estimated differential number flux $\hat{\Phi}$. (c)(d): Volume emission rate $\hat{P}$. (e)(f): Ground-observed intensity $B$ for characteristic energy $E_0 \in \{4.5, 1.6\}$ keV.

| $B_{\perp,0}$ [km] | $E_0$ [keV] | $\hat{B}_{\perp,0}$ [km] | $\hat{E}_0$ [keV] | Error $|B_{\perp,0} - \hat{B}_{\perp,0}|$ [%] | Error $|E_0 - \hat{E}_0|$ [%] |
|---|---|---|---|---|---|
| 1.0 | 4.5 | 1.0 | 4.1 | <5 | <10 |
| 1.0 | 1.6 | 1.0 | 1.7 | <5 | <10 |
| 1.55 | 5.0 | 1.55 | 4.67 | <5 | <10 |
| 2.5 | 4.5 | 2.5 | 4.1 | <5 | <20 |
| 2.5 | 1.6 | 2.55 | 1.2 | <5 | <25 |
| 3.75 | 5.0 | 3.7 | 5.7 | <5 | <25 |
| 4.2 | 4.5 | 4.15 | 5.6 | <5 | <25 |
| 4.2 | 1.6 | 4.1 | 1.15 | <5 | <30 |

TABLE I
SIMULATED ESTIMATION ERROR FOR FLAMING AND TRANSLATING AURORAL ARCS.

## REFERENCES

[1] C. Störmer, "Twenty-five years' work on the polar aurora," *Terrestrial Magnetism and Atmospheric Electricity*, vol. 35, no. 4, pp. 193–208, 1930.

[2] S. Frey, H. U. Frey, D. J. Carr, O. H. Bauer, and G. Haerendel, "Auroral emission profiles extracted from three-dimensionally reconstructed arcs," *Journal of Geophysical Research: Space Physics*, vol. 101, no. A10, pp. 21 731–21 741, 1996.

[3] R. A. Doe, J. D. Kelly, J. L. Semeter, and D. P. Steele, "Tomographic reconstruction of 630.0 nm emission structure for a polar cap arc," *Geophysical Research Letters*, vol. 24, no. 9, pp. 1119–1122, 1997.

[4] B. Gustavsson, "Tomographic inversion for alis noise and resolution," *Journal of Geophysical Research: Space Physics*, vol. 103, no. A11, pp. 26 621–26 632, 1998.

[5] J. Semeter, M. Mendillo, and J. Baumgardner, "Multispectral tomographic imaging of the midlatitude aurora," *Journal of Geophysical Research: Space Physics*, vol. 104, no. A11, pp. 24 565–24 585, 1999.

[6] C. C. Chaston, L. M. Peticolas, J. W. Bonnell, C. W. Carlson, R. E. Ergun, J. P. McFadden, and R. J. Strangeway, "Width and brightness of auroral arcs driven by inertial alfven waves," *Journal of Geophysical Research: Space Physics*, vol. 108, no. A2, 2003.

[7] J. P. McFadden, C. W. Carlson, and R. E. Ergun, "Microstructure of the auroral acceleration region as observed by fast," *Journal of Geophysical Research: Space Physics*, vol. 104, no. A7, pp. 14 453–14 480, 1999.

[8] J. Semeter, *Coherence in Auroral Fine Structure*, ser. Geophys. Monogr. Ser. Washington, DC: AGU, 2012, vol. 197, pp. 81–90.

[9] A. Jones, R. Gattinger, F. Creutzberg, F. Harris, A. McNamara, A. Yau, E. Llewellyn, D. Lummerzheim, M. Rees, I. McDade, and J. Margot, "The aries auroral modelling campaign: characterization and modelling of an evening auroral arc observed from a rocket and a ground-based line of meridian scanners," *Planetary and Space Science*, vol. 39, no. 12, pp. 1677 – 1705, 1991.

[10] H. Frey, S. Frey, B. Lanchester, and M. Kosch, "Optical tomography of the aurora and eiscat," *Annales Geophysicae*, vol. 16, no. 10, pp. 1332–1342, 1998.

[11] J. Semeter, M. Zettergren, M. Diaz, and S. Mende, "Wave dispersion and the discrete aurora: New constraints derived from high-speed imagery," *Journal of Geophysical Research: Space Physics*, vol. 113, no. A12, 2008.

[12] T. Pedersen, B. Gustavsson, E. Mishin, E. Kendall, T. Mills, H. C. Carlson, and A. L. Snyder, "Creation of artificial ionospheric layers using high-power hf waves," *Geophysical Research Letters*, vol. 37, no. 2, 2010, l02106.

[13] E. Kendall, R. Marshall, R. T. Parris, A. Bhatt, A. Coster, T. Pedersen, P. Bernhardt, and C. Selcher, "Decameter structure in heater-induced airglow at the high frequency active auroral research program facility," *Journal of Geophysical Research: Space Physics*, vol. 115, no. A8, 2010, a08306.

[14] J. E. Borovsky, "Auroral arc thicknesses as predicted by various theories," *Journal of Geophysical Research: Space Physics*, vol. 98, no. A4, pp. 6101–6138, 1993.

[15] J. Maggs and T. Davis, "Measurements of the thicknesses of auroral structures," *Planetary and Space Science*, vol. 16, no. 2, pp. 205 – 209, 1968.

[16] T. S. Trondsen, "High spatial and temporal resolution auroral imaging," Ph.D. dissertation, University of Tromsø, 1998. [Online]. Available: http://www.keoscientific.com/Documents/Trondsen_Dissertation_1998.pdf

[17] A. Omholt, *The Optical Aurora*, ser. Physics and Chemistry in Space. Springer-Verlag, 1971.

[18] H. Dahlgren, J. L. Semeter, R. A. Marshall, and M. Zettergren, "The optical manifestation of dispersive field-aligned bursts in auroral breakup arcs," *Journal of Geophysical Research: Space Physics*, vol. 118, no. 7, pp. 4572–4582, 2013.

[19] M.Hirsch, J.Semeter, M.Zettergren, H.Dahlgren, A.Baurley, C.Goenka, H.Akbari, and D.Hampton, "Multi-camera reconstruction of fine scale high speed auroral dynamics," Dec. 2014, poster SA13B-3991 presentated at AGU Fall Meeting.

[20] M. D. Zettergren, "Model-based optical and radar remote sensing of transport and composition in the auroral ionosphere," Ph.D. dissertation, Boston University, 2009. [Online]. Available: http://search.proquest.com/docview/304847517?accountid=9676

[21] P.-L. Blelly, A. Robineau, D. Lummerzheim, and J. Lilensten, "8-moment fluid models of the terrestrial high latitude ionosphere between 100 and 3000 km," in *Handbook of the Aeronomical Models of the Ionosphere*, B. Schunk, Ed. CASS, Utah State University, USA: Solar-Terrestrial Environment Program (STEP), 1996. [Online]. Available: http://scostep.apps01.yorku.ca/wp-content/uploads/2010/10/ionospheric-models.pdf

[22] S. M. Bailey, C. A. Barth, and S. C. Solomon, "A model of nitric oxide in the lower thermosphere," *Journal of Geophysical Research: Space Physics*, vol. 107, no. A8, pp. SIA 22–1–SIA 22–12, 2002.

[23] T. Sergienko and V. Ivanov, "A new approach to calculate the excitation of atmospheric gases by auroral electron impact," *Annales Geophysicae*, vol. 11, no. 8, pp. 717–727, Aug. 1993.

[24] D. J. Strickland, R. R. Meier, J. H. Hecht, and A. B. Christensen, "Deducing composition and incident electron spectra from ground-based auroral optical measurements: Theory and model results," *Journal of Geophysical Research: Space Physics*, vol. 94, no. A10, pp. 13 527–13 539, 1989.

[25] D. Lummerzheim and J. Lilensten, "Electron transport and energy degradation in the ionosphere: evaluation of the numerical solution, comparison with laboratory experiments and auroral observations," *Annales Geophysicae*, vol. 12, no. 10-11, pp. 1039–1051, 1994.

[26] C. C. Finlay, S. Maus, C. D. Beggan, T. N. Bondar, A. Chambodut, T. A. Chernova, A. Chulliat, V. P. Golovkov, B. Hamilton, M. Hamoudi, R. Holme, G. Hulot, W. Kuang, B. Langlais, V. Lesur, F. J. Lowes, H. Lühr, S. Macmillan, M. Mandea, S. McLean, C. Manoj, M. Menvielle, I. Michaelis, N. Olsen, J. Rauberg, M. Rother, T. J. Sabaka, A. Tangborn, L. Tøffner-Clausen, E. Thébault, A. W. P. Thomson, I. Wardinski, Z. Wei, and T. I. Zvereva, "International geomagnetic reference field: the eleventh generation," *Geophysical Journal International*, vol. 183, no. 3, pp. 1216–1230, 2010.

[27] K. Lancaster, "simplekml library," https://code.google.com/p/simplekml/, 2011–2015.

[28] D. Lang, D. W. Hogg, K. Mierle, M. Blanton, and S. Roweis, "Astrometry.net: Blind astrometric calibration of arbitrary astronomical images," *The Astronomical Journal*, vol. 139, no. 5, p. 1782, 2010. [Online]. Available: http://stacks.iop.org/1538-3881/139/i=5/a=1782

[29] J. Lilensten and P. Blelly, "The tec and f2 parameters as tracers of the ionosphere and thermosphere," *Journal of Atmospheric and Solar-Terrestrial Physics*, vol. 64, no. 7, pp. 775 – 793, 2002.

[30] M. Zettergren, J. Semeter, P.-L. Blelly, and M. Diaz, "Optical estimation of auroral ion upflow: Theory," *Journal of Geophysical Research: Space Physics*, vol. 112, no. A12, 2007.

[31] M. Zettergren, J. Semeter, P.-L. Blelly, G. Sivjee, I. Azeem, S. Mende, H. Gleisner, M. Diaz, and O. Witasse, "Optical estimation of auroral ion upflow: 2. a case study," *Journal of Geophysical Research: Space Physics*, vol. 113, no. A7, 2008.

[32] O. Tange, "Gnu parallel - the command-line power tool," *;login: The USENIX Magazine*, vol. 36, no. 1, pp. 42–47, Feb 2011. [Online]. Available: http://www.gnu.org/s/parallel

[33] Y.-M. Tanaka, T. Aso, B. Gustavsson, K. Tanabe, Y. Ogawa, A. Kadokura, H. Miyaoka, T. Sergienko, U. Brändström, and I. Sandahl, "Feasibility study on generalized-aurora computed tomography," *Annales Geophysicae*, vol. 29, no. 3, pp. 551–562, 2011.

[34] C. Simon Wedlund, H. Lamy, B. Gustavsson, T. Sergienko, and U. Brändström, "Estimating energy spectra of electron precipitation above auroral arcs from ground-based observations with radar and optics," *Journal of Geophysical Research: Space Physics*, vol. 118, no. 6, pp. 3672–3691, 2013.

[35] L. Peticolas and D. Lummerzheim, "Time-dependent transport of field-aligned bursts of electrons in flickering aurora," *Journal of Geophysical Research: Space Physics*, vol. 105, no. A6, pp. 12 895–12 906, 2000.

[36] D. W. Swift, "On the formation of auroral arcs and acceleration of auroral electrons," *Journal of Geophys. Res.*, vol. 80, pp. 2096–2108, 1975.

[37] F. Mozer and C. Kletzing, "Direct observation of large, quasi-static,

parallel electric fields in the auroral acceleration region," *Geophysical Research Letters*, vol. 25, pp. 1629–1632, 1998.

[38] R. E. Ergun, L. Andersson, D. S. Main, Y.-J. Su, C. W. Carlson, J. P. McFadden, and F. S. Mozer, "Parallel electric fields in the upward current region of the aurora: Indirect and direct observations," *Physics of Plasmas*, vol. 9, no. 9, pp. 3685–3694, 2002.

[39] D. J. Strickland, R. E. Daniell, J. R. Jasperse, and B. Basu, "Transport-theoretic model for the electron-proton-hydrogen atom aurora: 2. model results," *Journal of Geophysical Research: Space Physics*, vol. 98, no. A12, pp. 21533–21548, 1993.

[40] A. V. Jones, *Aurora*. D. Reidel Publishing Co., 1974.

[41] F. Kneizys, E. Shettle, L. Abreu, J. Chetwynd, G. Anderson, W. Gallery, J. Selby, and S. Clough, *Users guide to LOWTRAN 7*, Air Force Geophysics Laboratory, Hanscom AFB, Mass., 1988. [Online]. Available: http://www.dtic.mil/dtic/tr/fulltext/u2/a206773.pdf

[42] J. Semeter, "Ground-based tomography of atmospheric optical emissions," Ph.D. dissertation, Boston University, 1997. [Online]. Available: http://search.proquest.com/docview/304337205?accountid=9676

[43] W. M. Newman and R. F. Sproull, Eds., *Principles of Interactive Computer Graphics (2nd Ed.)*. New York, NY, USA: McGraw-Hill, Inc., 1979.

[44] J. Semeter and M. Mendillo, "A nonlinear optimization technique for ground-based atmospheric emission tomography," *Geoscience and Remote Sensing, IEEE Transactions on*, vol. 35, no. 5, pp. 1105–1116, Sep 1997.

[45] R. Byrd, P. Lu, J. Nocedal, and C. Zhu, "A limited memory algorithm for bound constrained optimization," *SIAM Journal on Scientific Computing*, vol. 16, no. 5, pp. 1190–1208, 1995.

[46] C. Zhu, R. H. Byrd, P. Lu, and J. Nocedal, "Algorithm 778: L-bfgs-b: Fortran subroutines for large-scale bound-constrained optimization," *ACM Trans. Math. Softw.*, vol. 23, no. 4, pp. 550–560, Dec. 1997.

[47] J. L. Morales and J. Nocedal, "Remark on "algorithm 778: L-bfgs-b: Fortran subroutines for large-scale bound constrained optimization"," *ACM Trans. Math. Softw.*, vol. 38, no. 1, pp. 7:1–7:4, Dec. 2011. [Online]. Available: http://doi.acm.org/10.1145/2049662.2049669

[48] E. Jones, T. Oliphant, P. Peterson *et al.*, "SciPy: Open source scientific tools for Python," 2001–, [Online; accessed 2014-11-06]. [Online]. Available: http://www.scipy.org/

[49] J. More and S. Wright, *Optimization Software Guide*, ser. Frontiers in Applied Mathematics. Society for Industrial and Applied Mathematics (SIAM), 1993, no. 14.

[50] A. Savitzky and M. J. E. Golay, "Smoothing and differentiation of data by simplified least squares procedures." *Analytical Chemistry*, vol. 36, no. 8, pp. 1627–1639, 1964.